\documentclass[twocolumn,aps,prd,   
               preprintnumbers,numbers,sort&compress,
               nofootinbib,
                            showpacs,
               colorlinks,
               linkcolor=blue,
               citecolor=blue]{revtex4-1} 

\usepackage{amssymb,graphicx}
\usepackage{amsmath}

\newcommand{\be}{\begin{equation}}
\newcommand{\ee}{\end{equation}}
\newcommand{\beq}{\begin{eqnarray}}
\newcommand{\eeq}{\end{eqnarray}}

\newcommand{\jcap}{JCAP}
\newcommand{\jgr}{Journal of Geophysical Research}

\newcommand{\aap}{Astronomy \& Astrophysics}
\newcommand{\mnras}{Mon.\ Not.\ R.\ Astron.\ Soc.}

\newcommand{\solphys}{Sol.\ Phys.}
\newcommand{\araa}{ARA \& A}
\newcommand{\physrep}{Physics Reports}
    \begin{document}
       \title{Solar Corona Heating by Axion Quark Nugget Dark Matter}
       \author{Nayyer Raza}
       \email{syednayyer.raza@alumni.ubc.ca}
        \affiliation{Department of Physics and Astronomy, University of British Columbia, Vancouver, V6T 1Z1, BC, Canada}
       \author{Ludovic Van Waerbeke}
       \email{waerbeke@phas.ubc.ca}
        \affiliation{Department of Physics and Astronomy, University of British Columbia, Vancouver, V6T 1Z1, BC, Canada}
       \author{Ariel  Zhitnitsky}
       \email{arz@phas.ubc.ca}
       \affiliation{Department of Physics and Astronomy, University of British Columbia, Vancouver, V6T 1Z1, BC, Canada}
     
       \begin{abstract}
       In this work we advocate for the idea that two seemingly unrelated 80-year-old mysteries - the nature of dark matter and the high temperature of the million degree solar corona - may have resolutions that lie within the same physical framework. The current paradigm is that the corona is heated by nanoflares, which were originally proposed as miniature versions of the observed solar flares. It was recently suggested that the nanoflares could be identified as annihilation events of the nuggets from the Axion Quark Nugget (AQN) dark matter model. This model was invented as an explanation of the observed ratio $\Omega_{\rm dark} \sim   \Omega_{\rm visible}$, based only on cosmological and particle physics considerations. In this new paradigm, the AQN particles moving through the coronal plasma and annihilating with normal matter can lead to the drastic change of temperatures seen in the Sun's Transition Region (TR), and significantly contribute to the extreme ultraviolet (EUV) excess of $10^{27}~{\rm erg~s^{-1}}$. To test this proposal, we perform numerical simulations with a realistically modeled AQN particle distribution and explore how the nuggets interact with the coronal plasma. Remarkably, our simulations predict the correct energy budget for the solar corona, and show that the energy injection mostly occurs at an altitude of around 2000 km, which is where the TR lies. 
  Therefore, we propose that these long unresolved mysteries could be two sides of the same coin. We make several predictions based on this proposal, some of which could be tested by the recently launched NASA mission, the Parker Solar Probe.
       \end{abstract}
     
       \maketitle

\section{Introduction}\label{sec:introduction}

Eighty years after the first evidence emerged supporting the existence of dark matter \cite{Zwicky}, its nature remains elusive despite numerous attempts at direct and indirect detections. For about 20 years, the standard paradigm for dark matter was based almost exclusively on the Weakly Interactive Massive Particles (WIMP), but the lack of detection prompted the development of alternative models. A promising approach was developed by Zhitnitsky \cite{Zhitnitsky:2002qa}, in the form of Axion Quark Nuggets (AQNs), where dark matter is, in part, composed of baryonic macroscopic objects (gram mass) and strongly interacting with the baryonic sector. At large (cosmological) scales, AQNs behave like cold dark matter because their high mass implies a low number density with a small cross-section. But at small scales, especially where the baryonic density is high, AQNs can interact strongly with baryons.

The idea that AQNs can take the form of composite objects of standard quarks in a novel phase, goes back to quark nuggets  \cite{Witten:1984rs}, strangelets \cite{Farhi:1984qu}, and nuclearities \cite{DeRujula:1984axn} (see also review \cite{Madsen:1998uh} which has a large number of references on the original results).  In the early models \citep{Witten:1984rs,Farhi:1984qu,DeRujula:1984axn,Madsen:1998uh}  the presence of strange quarks stabilizes the quark matter at sufficiently high densities, allowing strangelets being formed in the early universe to remain stable over cosmological timescales. Most of the original models were found to be inconsistent with some observations, but the AQN model was built on different ideas, involving the Axion field, and has not been ruled out so far. 
The AQNs could be made of matter as well as antimatter, where the latter would interact very strongly with baryons, and eventually annihilate, under certain conditions. We redirect the reader to Section \ref{sec:QNDM} for an introduction and overview of the basic ideas of the AQN model. In the same section we also highlight the basic cosmological and astrophysical consequences of this model. (For the interested reader we also refer to the short proceeding-type review   \cite{Lawson:2013bya} which has a large number of references on the original results obtained within the AQN framework.)

At a completely different scale and for different physics, the temperature of the solar corona is another 80-year-old puzzle: the photosphere is in thermal equilibrium at $\sim 5800$ K, while the corona has a temperature of a few $10^6$ K \cite{Grotrian-1939}. Observationally, the high temperature is seen as an energy excess of a few $10^{27}~{\rm erg~s^{-1}}$ and is mostly visible in the extreme ultraviolet (EUV) and soft X-ray regime. 

The conventional view is that the corona excess heating is supported by nanoflares, a concept originally invented by Parker \cite{Parker-1983}, which are thought to be miniature versions of the larger solar flares. The energy burst associated with these nanoflares is significantly below detection limits and has not been observed so far. Another approach is based on Alfv\'{e}n waves. In this type of model, energy is transported by Alfv\'{e}n waves from the photosphere up to the corona through the chromosphere. While there is still no consensus on which mechanism could be dominating the coronal heating, recent observations suggest that Alfv\'{e}n waves cannot provide a sufficient heat source \cite{Tomczyk-2007, Aschwanden-2000}. For the purposes of this paper, then, we direct our focus towards the currently accepted paradigm of nanoflares. In fact, all coronal heating models advocated so far seem to require the existence of an unobserved (i.e. unresolved with current instrumentation) source of energy distributed over the entire Sun \cite{De-Moortel-2015}. Therefore, `nanoflares' are largely modeled as generic events, producing an impulsive energy release on a small scale, without specifying their cause and their nature (see review papers \cite{Klimchuk:2005nx,Klimchuk:2017}). The exact nature of nanoflares, so far, remains an open question.


Our goal is to explore further the new assumption of \cite{Zhitnitsky:2017rop}, which is that the nanoflares (including sub-resolution events with very low energies) can be identified as the annihilation events of AQNs, thus providing an external and new source of energy to heat the corona. As pointed out in \cite{Zhitnitsky:2017rop,Zhitnitsky:2018mav}, the solar corona represents an ideal environment to test the interaction with AQNs: when AQNs enter a plasma under certain conditions, found in the solar corona, they can annihilate and deposit energy (the exact details of which are addressed later in the paper). The scenario as proposed by \cite{Zhitnitsky:2017rop,Zhitnitsky:2018mav} is that the AQN's annihilation provides an energy injection that can contribute to the observed EUV excess of the solar corona. In other words, the solar corona can play the role of a dark matter AQN detector, and provides a novel source of energy in the plasma. Since the AQN model predicts that annihilation, and energy release, will happen in the corona, the corona heating can serve as a robust test of the AQN model itself. This is the approach taken in this paper.



The presentation of the paper is organized as follows. First, in section \ref{sec:background} we introduce the basics of the solar corona physics, discussing the conventional approach to the heating problem and its limits, and expose our motivation for the present work. Then in section \ref{sec:QNDM} we overview the basic features of the AQN dark matter model, which is followed by section \ref{sec:aqn-nano} in which we develop the AQN model in the context of the solar corona environment. In section \ref{sec:numsim} we describe the setup for the numerical simulations performed to test our proposal and present our results. Concluding remarks, including possible future work, are addressed in section \ref{conclusion}.

\section{Background}\label{sec:background}

\begin{figure*}
\centering
\includegraphics[width=1.0\textwidth]{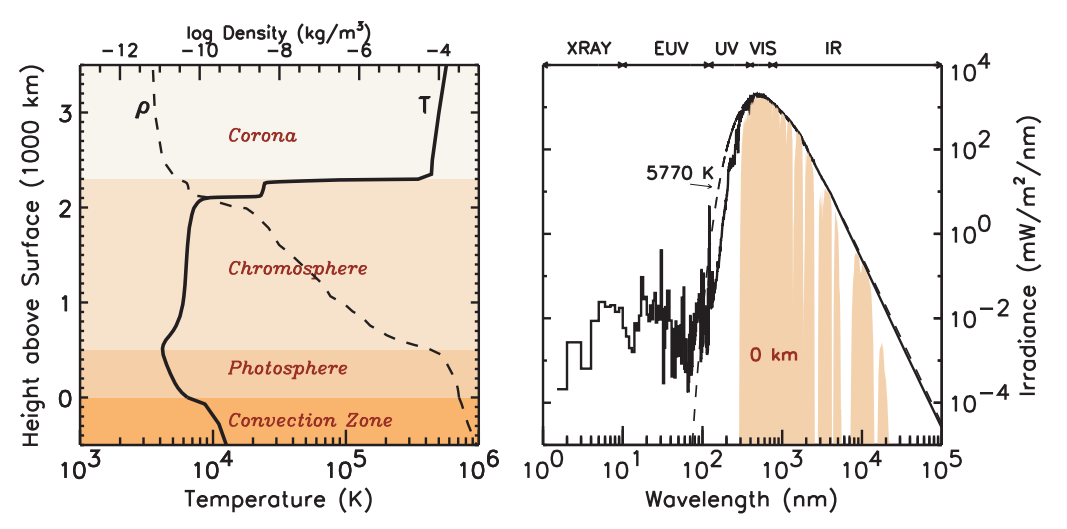}
\caption{\label{observations} Left: The temperature distribution of the inner and outer Sun. The drastic changes occur in vicinity of 2000 km.  Right: the unexpected deviation from the thermal distribution in the extreme ultraviolet (EUV) and soft X-rays in the solar spectrum constitutes the celebrated solar corona problem. This EUV and X-ray radiation is originated from chromosphere, transition and corona regions. The total EUV intensity  represents a  small $\sim (10^{-7}-10^{-6})$ portion of the solar irradiance. The plots are taken from \citep{Lean-1997}.}
\end{figure*}

\subsection{Physics of the Corona}

The solar corona is a very peculiar environment that seems to defy basic thermodynamics \citep{Law:2017}. Starting at an altitude of 1000 km above of the photosphere, the highly ionized iron lines show that the plasma temperature exceeds a few $10^6$ K. The total energy radiated away by the corona is of the order of $L_{\rm corona} \sim 10^{27} {\rm erg~ s^{-1}}$, which is about $10^{-6}-10^{-7}$ of the total energy radiated by the photosphere. Most of this energy is radiated at the extreme ultraviolet (EUV) and soft X-ray wavelengths. However, it is not in thermal equilibrium with its environment, since it is much hotter than the 5800 K blackbody temperature of the photosphere. As shown in Fig. \ref{observations}, there is a very sharp transition region located in the upper chromosphere where the temperature suddenly jumps from $\sim 25000$ K to $10^6$ K. This transition layer is relatively thin, 200 km at most. This apparent violation of the second principle of thermodynamics can only be resolved if there is a non-thermal source of energy, heating up the corona, located significantly above the photosphere. The source should be able to sustain a power of the order of $10^{27} {\rm ~erg~s^{-1}}$. It is important to note that the Sun can be approximately divided into active and quiet regions, and that, observationally, the EUV excess is found everywhere, in both regions. We want to emphasize that the problem we are discussing here concerns the {\it quiet} Sun, that is regions of the Sun away from active spots and coronal holes. The active regions give rise to powerful solar flares, but the energy injection provided by these spectacular events happens on a small area and have a negligible contribution to the overall heating of the corona. As we will explain in Sections \ref{sec:background}-B and \ref{sec:background}-C, it is unclear how conventional heating mechanisms can be efficient in the quiet Sun, where the magnetic field is small.



A conventional solution to the heating problem in the quiet Sun corona was proposed in 1983 by Parker \citep{Parker}, who postulated that a continuous and isotropic sequence of miniature flares, which he called ``nanoflares'', could happen in the corona.
Transient heating events, including `micro-events', `microflares' and `nanoflares', have been previously considered to be of potential interest for understanding the coronal heating mechanism  because  they may give rise to a basal background heating near the solar surface; see original papers \cite{Benz-2000,Benz-2001,Kraev-2001,Benz-2002,Benz-2003,Pauluhn:2006ut,Hannah:2007kw,Bingert:2012se} and reviews \cite{Klimchuk:2005nx,Klimchuk:2017,Jess-2014,Bradshaw-2012,terzo-2011,Kirichenko-2017,Cormack-2017}.
The term `nanoflare' has been used in a series of papers by Benz and coauthors
  \cite{Benz-2000,Benz-2001, Kraev-2001,Benz-2002, Benz-2003}, and many others, to advocate the idea that these small ``micro-events'' might be  responsible for the   heating of the   quiet solar corona.

According to \cite{Kraev-2001}, in order to reproduce the measured EUV excess, the observed range of nanoflares  needs to be extrapolated from the observed events  interpolating between $(3.1\cdot 10^{24}  - 1.3\cdot 10^{26})~{\rm erg}$ to sub-resolution events with much smaller energies of $\sim 10^{21}~{\rm erg}$.
The nanoflares have to be distributed very ``uniformly in quiet regions'', in contrast with micro-flares and flares
 which are much more energetic and occur exclusively in active areas \citep{Benz-2003}. 
 
 As we highlight in the next subsection, a conventional assumption that nanoflares and large flares (separated by many orders of magnitude in the energy scale) are originated from the same physics can be problematic for a number of reasons. Indeed, this was one of the major motivations to introduce an alternative description of the ``nanoflares" as the annihilating AQN dark matter particles.

\subsection{The conventional approach}

The conventional picture can be formulated  as follows:
 The flares and sunspots (which represent the direct and primary manifestation of the magnetic field activity) are strongly correlated spatially and temporally. It is normally assumed that the flare's energy is supplied by the magnetic reconnection events. This assumption is  supported by  Magneto-Hydro-Dynamics (MHD) simulations for sufficiently large magnetic fields\footnote{\label{MHD_1}To complicate the picture, the 2d MHD simulations \cite{Tanuma} show  that  a large number of different phenomena,
including Sweet-Parker  reconnection \cite{Sweet, Parker1}, Petschek reconnection \cite{Petschek, Kulsrud}, tearing instability, formation of the magnetic islands, and many others, may all take place at different phases in  the evolution of the system, see also reviews \cite{Shibata:2016,Loureiro}, but this is not very relevant for the discussion in this paper.}. 
Simulations are indeed consistent with observations when the magnetic field dynamics, observed by the sunspot activity, is correlated with the flare's activity, with solar cycles, and many other phenomena such as the Sun spots distribution over time (also known as the butterfly diagram), the emergence of the  coronal holes, and coronal mass ejections (CMEs). This is the case for flares in active regions where the magnetic field can be as high as $10^2-10^3$ G. Nanoflares, on the other hand, are expected to dominate the energy budget in the quiet regions, where the magnetic field is of the order of $\sim 1$ G, but occupies much larger solar surface area.


The conventional assumption is that the same physics of the magnetic reconnection (known to drive the large flares) can be extrapolated to much smaller scales such that the nanoflares have the same origin as large flares and are driven by the same physics of the magnetic reconnection.
This assumption is very hard to justify from theoretical, as well as from observational, viewpoints. To present our arguments in a quantitative way it is convenient to introduce the dimensionless plasma parameter $\beta$, which describes the ratio of gas pressure to magnetic pressure (and thus determines the importance of the magnetic field):
 \be
 \label{beta}
 \beta\equiv \frac{8\pi  p}{B^2} \sim 0.5\cdot  10^{3}\left(\frac{n}{10^{10} ~{\rm cm^{-3}}}\right) \left(\frac{T}{10^6 K}\right) \left(\frac{1 ~ G}{B}\right)^2,
 \ee
 where for numerical estimates we use typical parameters for the quiet  regions in corona when $\beta\gg 1$.
 Another important parameter is the Alfv\'{e}n speed $v_A$ which assumes the following numerical value in the corona environment:
 \be
 \label{alfven}
 \frac{v_A}{c}=\frac{B}{c\sqrt{4\pi\rho}} \sim 2\cdot 10^{-5}\left(\frac{B}{1~ G}\right)\cdot \sqrt{ \frac{10^{10} ~{\rm cm^{-3}}}{n} },
 \ee

 The large value of parameter (\ref{beta}) implies that the magnetic  field pressure plays a subdominant role
 in comparison with conventional kinetic pressure $p$. It  is very hard to see how the magnetic reconnection
 could be operational in the environment  when the magnetic pressure  is 3 orders of magnitude smaller than conventional kinetic pressure\footnote{In most cases the MHD simulations are done with a small plasma parameter $\beta\leq1$, i.e. when the magnetic field dynamics dominate the physics.}. The
 Alfv\'{e}n speed in this environment is also numerically very   small: $ v_A\simeq 6~ {\rm km\,s^{-1}}$.

The assumption that flares and nanoflares are similar phenomena is also difficult to justify observationally, due to the following:

{\bf 1.} Flares have a highly non-isotropic spatial distribution because they are associated with the active regions. On the other hand, the EUV emission is highly isotropic. In order to explain the SoHo/EIT observations, a large rate of $1.1\times 10^6$ events per hour for whole Sun is necessary \citep{Benz-2001, Benz-2002}. This large number of events is required to fit the observations when the
  EUV iron lines fluctuate locally at time scales of a few minutes in a majority of pixels, including even the intra-cell regions of the quiet corona.

{\bf 2.} The nanoflares and microflares appear in different ranges of temperature and emission measure (see  Fig.3 in \cite{Benz-2003}). While  the instrumental  limits prohibit observations at intermediate temperatures, nevertheless the authors of \cite{Benz-2003} argue that  ``the occurrence rates of nanoflares and microflares are so different that they cannot originate from the same population". We emphasize on this difference to argue that the flares originate at sunspot areas with locally large magnetic fields $B\sim (10^2-10^3)$ G, while  the EUV emission (which is observed  even in very quiet regions where the magnetic  field  is in the range $B\sim 1$G) is isotropic and covers the entire solar surface.

{\bf 3.} The temporal evolution of flares and nanoflares also appears different. The typical ratio between the maximum and minimum EUV irradiance during  the solar cycle does not exceed a factor of 3 or so between the maximum at year 2000 and minimum in 2009 (see Fig. 1 from ref. \cite{Bertolucci-2017}), while the same ratio for flares and sunspots is much larger, of the order of $10^2$.

If the magnetic reconnection was fully responsible for both the flares and nanoflares, then  the variation during the solar cycles should be similar for these two phenomena. It is not what is observed; the modest variation of the EUV with the solar cycles in comparison to the flare fluctuations suggests  that the EUV radiation does not directly follow the magnetic field activity, and that the EUV fluctuation is a  secondary, not a primary effect of the magnetic activity.

\subsection{Motivation and contribution to the conventional approach}

 In the present work, we advocate an alternative idea: the source and mechanism behind `nanoflares' lies in different physics compared to large flares. Following a proposal by \cite{Zhitnitsky:2017rop,Zhitnitsky:2018mav}, nanoflares are associated with AQN annihilation events, and are not related to the solar  magnetic field and accompanying magnetic reconnection. This model provides an external source of energy to the solar corona.
Since the AQN model was initially designed to address cosmological issues only, 
its development has no connection with solar physics and its interaction with the solar corona cannot be tuned; it is a direct consequence of the initial model.
If dark matter, in the form of AQNs, can heat the solar corona, the energy available in the solar system dark matter environment should be a reasonable estimate of the energy injected in the corona. Surprisingly, with a dark matter mass density of $\rho_{DM} \simeq 0.3 \ {\rm GeV\,cm^{-3}}$ the power potentially available for the corona is of the order of $10^{27}~ {\rm erg~s^{-1}}$, which is very close  to the observed EUV excess.

Compared to the conventional approach, the only new element in our proposal is related to the nature of the nanoflares: in the AQN framework these nanoflares are not expressed in terms of conventional solar  physics, and cannot be described in terms of the magnetic reconnection. However, all other phenomena, such as the statistics of large flares, their spatial and temporal correlation with sunspot activity, CMEs, and variations with the solar cycle and magnetic activity remains unaffected. Our assumption on the nature of nanoflares is also consistent with MHD simulations, where ``nanoflares" are treated as generic energy burst events, without specifying their cause and nature (see review papers \cite{Klimchuk:2005nx,Klimchuk:2017}).

     
The time variability and spatial distribution of nanoflares are also important clues about the nature of non-thermal processes happening in the chromosphere. Recent RHESSI observations demonstrate clearly that nanoflares and microflares are different physical phenomena \cite{Benz-2003}. Microflares are well resolved and similar to a miniature version of the solar flares, appearing preferentially in active regions, with a higher temperature and emission measure. Nanoflares however tend to appear uniformly on the Sun and have very distinct energetics compared to microflares.

\section{The Axion Quark Nugget (AQN) dark matter model}\label{sec:QNDM}

The AQN model in the title of this section stands for the axion quark nugget   model, see original work \cite{Zhitnitsky:2002qa} and  short overview
\cite{Lawson:2013bya}  with references therein on the original results reflecting different aspects of the AQN model.


  The original motivation of this model was based on the observation  that the visible and dark matter densities in the Universe are of the same order of magnitude \citep{Zhitnitsky:2002qa} . Indeed, this order of magnitude equality is automatically realized in the AQN model
\be
\label{Omega}
 \Omega_{\rm dark}\sim \Omega_{\rm visible}
\ee
as both densities are proportional to the same fundamental $\Lambda_{\rm QCD} $ scale,
and they both originate from the same  QCD epoch, see \cite{Liang:2016tqc,Ge:2017ttc,Ge:2017idw} with many technical  details. If these processes are not fundamentally related, the two components $\Omega_{\rm dark}$ and $\Omega_{\rm visible}$ could easily exist at vastly different scales; this is a fine tuning problem which is rarely discussed in the literature.

In comparison with many other similar proposals \cite{Witten:1984rs,Farhi:1984qu,DeRujula:1984axn,Madsen:1998uh}, the AQN dark matter model  has  two unique  features:\\
1. There is an  additional stabilization factor in the AQN  model provided    by the {\it axion domain walls}
  which are copiously produced during the QCD transition in early Universe;\\
  2. The AQNs  could be
made of matter as well as {\it antimatter} in this framework as a result of separation of the baryon charges.

 The most  important astrophysical implication  of these new aspects   relevant for the present studies    is that quark nuggets made of  antimatter
 store a huge amount of energy which can be released when the anti-nuggets hit the Sun from outer space  and get annihilated.  This feature
 of the AQN model is unique and is not shared by any other dark matter models because the dark matter in AQN model is made  of the same quarks and antiquarks of the standard model (SM) of particle physics \footnote{In general, the   annihilation events of the anti-nuggets with visible matter  may  produce a number of other observable effects in different  circumstances such as  rare events of annihilation of anti-nuggets with  visible matter in the center of the galaxy, or in the Earth's atmosphere (see some  references on the original computations for different frequency bands in the short review \cite{Lawson:2013bya})}.

The basic idea of  the AQN  model can be summarized   as follows:
It is commonly  assumed that the Universe
began in a symmetric state with zero global baryonic charge
and later (through some baryon number violating process, the so-called baryogenesis)
evolved into a state with a net positive baryon number. As an
alternative to this scenario we advocate a model in which
``baryogenesis'' is actually a charge separation process
when  the global baryon number of the Universe remains
zero. In this model the unobserved antibaryons come to comprise
the dark matter in the form of dense nuggets of quarks and antiquarks in the color superconducting (CS) phase.
  The formation of the  nuggets made of
matter and antimatter occurs through the dynamics of shrinking axion domain walls (see original papers \cite{Liang:2016tqc,Ge:2017ttc,Ge:2017idw} which contain many technical  details).

The nuggets, after they are formed, can be
viewed as strongly interacting and macroscopi-
cally large objects, with a typical nuclear den-
sity and with a typical size $R\sim (10^{-5}-10^{-4})$ cm determined by the axion mass $m_a$, as these two parameters are linked: $R\sim m_a^{-1}$. The most strin-
gent upper bound on the axion mass comes from
the observation of supernova SN 1987A, while the
lower bound is determined by the requirement
that the energy density of axion dark matter does
not over-close the universe. We refer the reader
to the recent reviews \cite{vanBibber:2006rb, Asztalos:2006kz,Sikivie:2008,Raffelt:2006cw,Sikivie:2009fv,Rosenberg:2015kxa,Marsh:2015xka,Graham:2015ouw,Ringwald:2016yge} on the subject. For the purposes of the present work it is
sufficient to mention that the conventional dark
matter axions in the galaxy are produced due to
the misalignment mechanism or due to the decay
of the topological objects. In the corresponding
computations it has also been assumed that the
Peccei-Quinn symmetry was broken after inflation.

Taking these factors into account, the remaining open window for the axion mass is then $10^{-6} {\rm eV}\leq m_a \leq 10^{-2} {\rm eV}$. This axion mass window corresponds to the range of the nugget's baryon charge $B$:
\be
\label{B-range}
10^{23}\leq |B|\leq 10^{28}, ~~~{\cal{M}}\sim m_pB
\ee
where $\cal{M}$ is the  mass  of the nugget and  $m_p$ is the proton mass. One should emphasize that while the two parameters, the nugget's size $R$ and the axion mass $m_a$ are linked as mentioned above, this relation is not one to one correspondence. To be  more specific,
  for a given axion mass $m_a$ there is entire window
 for the  baryon charge $B$ where the nuggets remain stable as the quark energy per baryon charge in the CS phase is still below than $m_p$, which represents the energy per baryon charge in the hadronic phase, see (\cite{Ge:2017idw} for details).  
Therefore as the  baryon charge scales as $B\sim R^3$ while  $R$ could   easily vary by a factor of 3-4 depending on the QCD model (see e.g. Fig 8 from  \cite{Ge:2017idw}), we expect the baryon charge of the nuggets to be
distributed in a relatively large window covering
a few orders of magnitude.

The corresponding high mass of the nuggets implies a very small number density $\sim B^{-1}$. As a result, their interaction  with visible matter is highly  inefficient, and the nuggets behave as cold dark matter. Therefore, the AQN model does not contradict any of the many known observational constraints on dark matter or
antimatter  in the Universe \citep{Zhitnitsky:2006vt}.


Furthermore, it is known  that the galactic spectrum contains several excesses of diffuse emission of uncertain origin, the best known example being the strong galactic 511~keV line. If the nuggets have the  average  baryon number in the $\langle B\rangle \sim 10^{25}$ range they could offer a potential explanation for several of these diffuse components (including the 511 keV line and accompanied  continuum of $\gamma$ rays in the 100 keV to few  MeV ranges, as well as X-ray and radio frequency bands). For further details see the original works \cite{Oaknin:2004mn, Zhitnitsky:2006tu,Forbes:2006ba, Lawson:2007kp, Forbes:2008uf,Forbes:2009wg}   with specific computations in different frequency bands in galactic radiation, and a short overview \cite{Lawson:2013bya}.



\section{AQN in the Solar Corona}\label{sec:aqn-nano}

\subsection{The AQN Annihilation Events as Nanoflares}

\subsubsection{Energetics}

We  want to  overview here the basic results of \cite{Zhitnitsky:2017rop} suggesting that the heating of the chromosphere and corona is due to the annihilation events of the AQN with the solar material. Indeed, the impact parameter for capture of the nuggets by the Sun can be estimated as follows:
  \be
  \label{capture}
  b_{\rm cap}\simeq R_{\odot}\sqrt{1+\gamma_{\odot}}, ~~~~ \gamma_{\odot}\equiv \frac{2GM_{\odot}}{R_{\odot}v^2},
  \ee
  where $v\simeq 10^{-3}c$ is a typical velocity of the nuggets. Assuming that $\rho_{\rm DM} \simeq 0.3~ {\rm GeV\,cm^{-3}}$ and using the capture impact parameter (\ref{capture}), one can estimate
  the total energy flux due to the complete annihilation of the nuggets,

  \be
  \label{total_power}
   L_{\odot ~  \rm (AQN)}\sim 4\pi b^2_{\rm cap}\, v\, \rho_{\rm DM}
  \simeq 4.8 \cdot 10^{27} {\rm erg~s^{-1}},
  \ee
   where we substitute a constant $v\simeq 10^{-3}c$  to simplify numerical  analysis.
There is a non-trivial coincidence between this estimate and the observed total EUV  energy output from the corona. As highlighted in Section II, it is hard to explain the EUV excess in terms of conventional astrophysical sources. This ``accidental  numerical coincidence" was the main motivation   to put forward the idea that  (\ref{total_power}) represents a new source of energy feeding the EUV and soft X-ray radiation \citep{Zhitnitsky:2017rop}.


   One should emphasize that the estimates (\ref{total_power})   for the radiated power as well as
   the estimate   for a typical temperature  $T\sim 10^6$ K
   are not very sensitive to the size distribution of the nuggets. This is because  the  estimate (\ref{total_power})   represents the total energy input due to the complete nugget's annihilation, while their total baryon charge is determined by the dark matter density  $\rho_{\rm DM} \sim 0.3~ {\rm GeV\,cm^{-3}}$
   surrounding the Sun.

\subsubsection{Energy distribution}

The expected energy distribution of nanoflares also overlaps with the baryon charge distribution of AQNs. The energy distribution derived from studying models of coronal heating by nanoflares (see e.g. \cite{Kraev-2001}), is a power-law formally expressed as
  \begin{eqnarray}
  \label{distribution}
   {dN} & \sim & W^{-\alpha_{\rm nano}}dW\sim B^{-\alpha_{\rm nano}} dB, \nonumber \\
   \rm for ~~~~ W & \simeq &   (4\cdot 10^{20}  -  10^{26})~{\rm erg}
  \end{eqnarray}
  where $dN$ is the number of the nanoflares  (including the sub-resolution events) per unit time with energy between $W$ and $W+dW$. By identifying these nanoflare events with annihilation events of the AQN carrying  the  baryon charges between  $B$ and $B+dB$, the two distributions become tightly linked in our framework. More concretely, as the annihilation of a single baryon charge deposits an energy of $2 m_pc^2$ into the corona, the energy of the events $W$ can always be expressed in terms of the baryon charges $B$ of the AQNs:
  \be
 \label{eq:aqn-nano}
 W\simeq 2 m_pc^2 B \ \simeq (3\cdot 10^{-3} \ {\rm erg}) \times B
 \ee
  
One can see that the nanoflares energy distribution window given by eq. (\ref{distribution}) largely overlaps with the AQN baryonic charge window given by eq. (\ref{B-range}). One should emphasize that this overlap is a nontrivial self-consistency check of our proposal connecting nanoflares to AQNs, since the nanoflare window (\ref{distribution}) is constrained by solar corona heating models, while the nugget's baryon charge  window  (\ref{B-range}) is constrained by cosmological, astrophysical, satellite and ground based observations and experiments, including the axion search experiments.
  
  The following comment will also be useful for the rest of the paper: the authors of \cite{Pauluhn:2006ut} claim that the the data prefer a nanoflare energy distribution  (\ref{distribution}) with a slope $\alpha_{\rm nano}\simeq 2.5$, while numerous attempts to reproduce the data with $\alpha_{\rm nano} < 2$ were unsuccessful. This is consistent with previous analysis
 \cite{Benz-2002} with $\alpha_{\rm nano}\simeq 2.3$.
It should be contrasted with another analysis \cite{Bingert:2012se} which suggests that $\alpha_{\rm nano}\simeq 1.2$ for events below $W \leq 10^{24}$ erg,   and $\alpha_{\rm nano}\simeq 2.5$ for events above $W \geq 10^{24}$ erg.  Analysis \cite{Bingert:2012se}  also suggests that the change of the scaling (the position of the knee) occurs at energies close to $\langle W\rangle \simeq 10^{24}~ {\rm erg}$, which roughly coincides with the maximum of the energy distribution, see figure 7 in \cite{Bingert:2012se}.

\subsubsection{Dynamics}

The last aspect that could potentially link nanoflares to AQNs is dynamical. Observations of lines in the solar corona reveal large Doppler shifts, with typical velocities of $(250-310)~ {\rm km\,s^{-1}}$ (see figure 5 in \cite{Benz-2000}).  The observed line width in OV  of $\pm 140~ {\rm km\,s^{-1}}$ far exceeds the thermal ion velocity which is around $11~ {\rm km\,s^{-1}}$   \cite{Benz-2000}. On the other hand, it is comparable to the typical velocities of the nuggets entering the solar corona which is of the order of $ \sim  300~ {\rm km\,s^{-1}}$. Typical timescales of the nanoflare events, of the order $10^1-10^2$ seconds, are also consistent with with AQN annihilation estimates \cite{Zhitnitsky:2017rop}. Both quantities, velocity and timescale, will be more precisely calculated in Section V.

One should add that the observations listed in items {\bf 1, 2, 3} from Section \ref{sec:background}-B find a natural explanation within the AQN framework. Indeed, according to these items the nanoflares are distributed very uniformly in quiet regions, in contrast with micro-flares which are much more energetic and occur exclusively in active areas. This is consistent with the dark matter interpretation as the  anti-nugget annihilation events (identified with nanoflares) should be present in all areas irrespective of the regional activity in the Sun. The same anti-nugget annihilation events also occur during low solar activity periods when no active regions or flares are present in the system for months. It is consistent with the observations that the EUV intensity  fluctuations (which according to this proposal are due to the AQN annihilation events) are very modest in comparison with the drastic changes of flare activity during a solar cycle.
 
 

\subsection{Formulation of the Interaction Cross-Section}

In this section we highlight and further develop the basic ideas from  \cite{Zhitnitsky:2017rop} with estimations of the rate of ionization  of the nuggets (and antinuggets) as a result of their high speeds in the corona. The corresponding estimates will play a central role in the numerical calculations developed in Section V.

We start with the estimation of  the electrical charge of the AQNs when they enter the solar corona. The total neutrality of the nuggets in the model is supported by the electrosphere made of leptons (electrons for nuggets and the positrons for the anti-nuggets). For a non-zero intrinsic nugget temperature $T\neq 0$   a small portion of the loose positrons will be stripped off from the AQNs, such that
 the nuggets will be ionized at $T\neq 0$.
 As a result the nuggets will acquire a non-vanishing positive charge, while  anti-nuggets will  acquire a non vanishing negative electric charge $Q$. To estimate this charge $Q$ one can use the electro-sphere density profile function $n(r)$ by removing the contribution of the region of loosely bounded  positrons with  low momentum $p^2\leq 2 m_e T$.   The corresponding computation leads to the following estimate  for $Q$ (see  \cite{Zhitnitsky:2017rop}):
  \be
  \label{Q}
Q\simeq 4\pi R^2 \int^{\infty}_{\frac{1}{\sqrt{2 m_e T}}}  n(z)dz\sim \frac{4\pi R^2}{2\pi\alpha}\cdot \left(T\sqrt{2 m_e T}\right).~~~~~
  \ee
  If we assume a typical AQN size  $R\sim 10^{-5} \ {\rm cm}$ and $T\sim 100$ eV corresponding to the temperature of the surrounding plasma in the corona we arrive at the estimate  $Q\sim 10^{8}$, which represents a very small portion in comparison with the typical baryon charge $B\sim 10^{25}$ hidden in the AQNs, i.e. $(Q/B)\ll 1$.  One should emphasize that our estimate $T\sim 100$ eV is actually a lower limit for an estimation of the charge $Q$, because the corresponding temperature entering eq. (\ref{Q}) should be identified with the internal thermal temperature $T_I$ of the nuggets (and anti-nuggets), to be contrasted  with the  surrounding plasma temperature $T_P$ measured far away from the nuggets. The $T_I$ could be many orders of magnitude higher than the average plasma temperature $T_P\sim 100$ eV, and so the charge $Q \sim T_I^{3/2}$ could also be drastically different in magnitude. It is worth noting that we could also expect the internal local temperatures for the nuggets versus the antinuggets to be drastically different, because heating from the proposed annihilation events occur exclusively inside the antinuggets, while the nuggets are heated exclusively as a result of the supersonic motion in the surrounding plasma. The estimates and arguments for the effective cross section formulated in the following paragraphs, which attempt to account for the internal and plasma temperature differences $T_I$ and $T_P$ due to supersonic motion, are then only lower limits for the anitnuggets (for they do not account for the annihilation heating).

  In our numerical simulations that follow, it is important that we have formulations for the calculation of the effective interaction sizes of the nuggets and antinuggets.
  The simplest and very rough way to estimate the corresponding parameter $R_{\rm eff}$ (effective radius of the spherical AQN) is to approximate an effective Coulomb cross section between the nuggets carrying the charge $Q$ and the plasma of the electrons and protons by assuming that a typical momentum transfer is order of the temperature of the surrounding plasma, $|q|\sim T_P$, i.e.
 \be
 \label{momentum}
 \pi R_{\rm eff}^2\sim \frac{Q^2\alpha^2}{q^2}\sim \frac{Q^2\alpha^2}{T_P^2}.
 \ee
 So now we can estimate $ R_{\rm eff}$ using the ionization  charges determined by eq. (\ref{Q}):
 \be
 \label{T}
  \left(\frac{R_{\rm eff}}{R}\right)^2\simeq \frac{8 (m_e T_P) R^2}{\pi}\left(\frac{T_I}{T_P}\right)^3
 \ee
 Or equivalently, we define for the purposes of our simulations:
 \be
 \label{eps1}
  \left(\frac{R_{\rm eff}}{R}\right) = \epsilon_1 \left(\frac{T_I}{T_P}\right)^{3/2}, \qquad \epsilon_1 \equiv \sqrt[]{\frac{8 (m_e T_P) R^2}{\pi}}
 \ee
 Where $\epsilon_1$ is defined to be understood as a dimensionless enhancement factor for the nugget interaction radius. If we ignore the difference between the temperatures $T_I$ and $T_P$ we arrive at an estimate for $R_{\rm eff}$ (from $\epsilon_1$, which then effectively determines the size of the system) as
 \begin{eqnarray}
 \label{R_eff}
 \left(\frac{R_{\rm eff}}{R}\right) & \simeq &  10^4~~ \Rightarrow  ~~ R_{\rm eff}\sim 0.1~ {\rm cm} \nonumber \\
  {\rm for} ~~Q & \sim &  10^8~~ {\rm and}~~  T_P\sim 10^6 \ {\rm K}.
 \end{eqnarray}
  Precisely this value  $ R_{\rm eff}\sim 0.1~ {\rm cm}$ has been used in an order of magnitude estimate in \cite{Zhitnitsky:2017rop}.

  The effective radius $R_{\rm eff}$  of the AQNs can be interpreted as an effective size of the nuggets due to the ionization characterized by  the nugget's charge $Q$. It can also be thought of as a typical  radius  of a sphere  which can accommodate $\sim n_{\rm sun}(l)R^3_{\rm eff}(l)$  number of particles from plasma. Precisely these particles effectively participate in the processes of annihilation and energy transfer from the antinugget to the surrounding solar plasma. The corresponding value of $R_{\rm eff}(l)$ obviously depends on the environmental parameters such as density $n_{\rm sun}(l)$ and the temperature $T_P(l)$ of the plasma. This feature is reflected by dependence of the internal temperature on the altitude $l$.

To account for the physics related to the   difference between internal temperature $T_I$ and plasma temperature $T_P$ we first define the corresponding dimensionless parameter:
    \be
 \label{epsilon}
 \epsilon_2\equiv \left(\frac{T_I}{T_P}\right)^{3/2} \quad \Rightarrow \quad \left(\frac{R_{\rm eff}}{R}\right) = \epsilon_1 \epsilon_2
 \ee
So that in what follows $\epsilon_1$ and $\epsilon_2$ are treated as the phenomenological enhancement parameters.

An estimation of the internal thermal temperature $T_I$ (or what is the same $\epsilon_2$) is a highly nontrivial and complicated problem and requires an understanding of how the heat (due to the friction and the annihilation events continuously occurring inside the antinuggets)  will be transferred to the surrounding plasma from a body moving with supersonic speed with Mach number $M\equiv v/c_s>1$. The efficiency of this heat transfer eventually determines the internal thermal temperature of a nugget and the corresponding charge $Q$.
 The corresponding energy transfer efficiency depends on the number of many body plasma phenomena, including turbulence in the vicinity of the nugget's surface.
 Such an estimate of the internal temperature $T_I$ is well beyond the scope of the present work. As we mentioned, in what follows we treat $\epsilon_2$ as a phenomenological parameter. However, one could get a rough estimate on the magnitude of $T_I$ using simple thermodynamical arguments which go as follows.

  It has been argued in \cite{Zhitnitsky:2018mav} that the nuggets in the corona will inevitably generate shock waves due to their very large Mach number, which was estimated as $M\simeq (1.5-15)$ depending  on the typical velocities of the nuggets. It is known that a shock wave generates a discontinuity in temperature, which for large Mach numbers $M\gg 1$ can be approximated as follows \cite{Zhitnitsky:2018mav, Landau}
  \be
 \label{shock1}
  \frac{T_2}{T_1}\simeq M^2\cdot \frac{2\gamma(\gamma-1) }{(\gamma+1)^2}, ~~~ \gamma\simeq 5/3.
 \ee
 In this formula  we identify the  temperature $T_1\simeq T_P$  with the temperature of the surrounding unperturbed plasma, while the high temperature $T_2$ occurs as a result of the shock wave. If one assumes that the turbulence (which normally develops around a body moving with supersonic speed) will efficiently equalize the internal temperature of the nuggets $T_I$ with $T_2$ one can estimate from eq. (\ref{shock1}) that $T_I/T_P\sim M^2$, which could be very large as the factor $M\sim 10$ could be very large. This effect obviously applies to both types of the AQNs: nuggets and antinuggets. However, we note again that we expect $T_I$ for the antinuggets could actually be much larger than $T_I$ for the nuggets once the antinuggets start to annihilate in the Sun and have an additional internal heat generated as a result. In any case, these estimates suggest that $\epsilon_2$ could be numerically very large as it scales with the internal temperature as $T^{3/2}$. The important result for our work, then, is that the parameter $\epsilon_2$ scales with and is determined by the Mach number as follows:
 \be
 \label{epsilon1}
 \epsilon_2\equiv \left(\frac{T_I}{T_P}\right)^{3/2}\sim M^3.
 \ee
 Therefore, the parameters $\epsilon_1$ and $\epsilon_2$ depend on altitude as well as on the AQN velocity at each given point, as the AQN velocity obviously changes with time as a result of friction and annihilation events. The corresponding modifications of the parameters $\epsilon_1$ and $\epsilon_2$ when time evolves, and thus the evolution of the effective interaction cross section, will be explicitly accounted for in our numerical studies in the next section.

\section{Numerical simulations}\label{sec:numsim}

The main aim of this work was to investigate the feasibility and accuracy of the proposed model of the AQN dark matter particles as a source of the heating of the corona through nanoflare-type events. To do this, we performed detailed numerical simulations of the entire proposed process, paying particular attention to the solar environment. We divided our simulations into three main steps: in the first step we generated the dark matter particles in the solar neighborhood and calculated their trajectories, in the second step we identified these particles as AQN and assigned masses to them, and in the third step we solved the equations for annihilation of these AQN in the solar atmosphere.

\subsection{Numerical Setup}

\subsubsection{\label{1}DM particles in the solar neighborhood}

For the initial set-up, we first populated the solar neighborhood with a large sample of particles with randomly assigned positions and velocities from known probability distributions, i.e. a Monte Carlo sampling. The rotational velocity of the Sun relative to the galactic center is $V_c \simeq 220 \ {\rm km\,s^{-1}}$. We also assume that the dark matter halo is not rotating relative to the galactic center; \cite{Bett-2010} showed that the halo rotation speed is of the order of $10~ {\rm km\,s^{-1}}$, which is negligible compared to $V_c$. In the halo frame, the dark matter particles follow a NFW density profile with an isotropic velocity distribution given by a three dimensional Maxwellian distribution. The velocity dispersion per component $\sigma_{v_i}$ must be calculated from the Jeans equation, at the Sun's location, which is $0.04 r_{\rm vir}$, where $r_{\rm vir}\simeq 200~{\rm kpc}$ is the virial radius of the Milky Way. Considering a Milky Way mass of approximately $10^{12} ~ M_{\odot}$, the velocity dispersion per component is $\sigma_{v_i} \simeq 100~{\rm km\,s^{-1}}$ at the Sun's location \cite{Lokas-Mamon-2001,Dehnen-2006}, where we have assumed a spherical dark matter halo.
Consequently, the full velocity distribution of AQN particles is given by a three dimensional Maxwellian distribution shifted in one direction, given by the equation

\be
\label{velocity_dist}
f_{\bf v}(v_x, v_y, v_z) = \frac{1}{\sqrt[]{2\pi \sigma_{v_i}}}  \ exp\left[-\frac{(v_x-v_{\odot})^2 + v_y^2 + v_z^2}{2\sigma_{v_i}^2}\right].
\ee
The positions of the particles are such that a spherical annulus of radii $R_{\rm max} = 10 \ AU, R_{\rm min} = R_{\odot}$ around the Sun is populated uniformly (i.e. the probability of finding a particle in a volume element $dV$ is constant throughout the entire volume). To generate the uniform distribution of particle positions, we used the following coordinate equations:
\beq
\label{uniform_sphere_positions}
r &=& \Big[\left(R_{\rm max}^3 - R_{\rm min}^3\right) u + R_{\rm min}^3 \Big]^{1/3} \\
\theta &=& \cos^{-1}\left(2v - 1\right) , ~~~
\phi = 2 \pi w  , ~~~~
u, v, w \sim {\rm Unif}(0, 1)  \nonumber
\eeq
We can then generate the Monte Carlo sampled 3D positions and velocities for each particle. We generated $2 \times 10^{10}$ such sample particles, and let them move according to Newton's law of gravity. Note that this number is not the true number of DM particles that exist in the solar neighborhood, but only a small representative fraction, chosen due to computational limitations. A rescaling procedure to match the actual number density of dark matter particles will be given in Section 6.2.

\begin{figure*}
\centering
\includegraphics[width=0.9\textwidth]{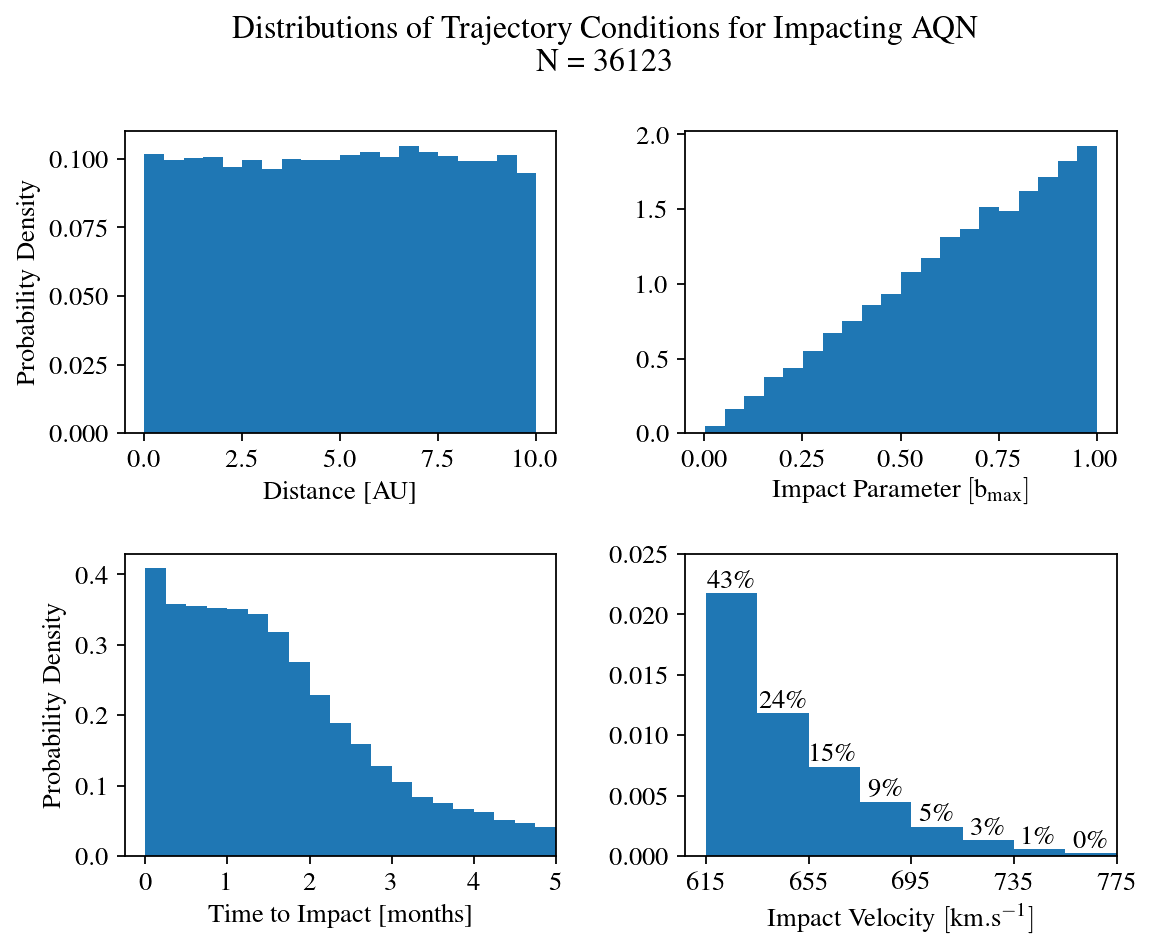}
\caption{\label{AQN_initial_dists} Probability density distributions of the trajectory conditions for the 36,123 impacting AQN dark matter particles. It can be seen that the probability of impact is distance independent, while the impact parameter scales linearly, where $b_{\rm max} \equiv R_{\odot} \sqrt{1 + \left(2GM_{\odot}/R_{\odot}v_{i}^2\right)}$ ($v_i$ being the initial velocity drawn from eq. (\ref{velocity_dist})). The window of time approximately between 0.25 and 1.25 months, where the `time to impact' distribution is constant (i.e. the AQN flux becomes constant), is used to extrapolate the total impact rate and the total luminosity from those impacts.}
\end{figure*}

Once we have the position and velocity for each particle, we calculated the trajectory for each particle using classical two-body orbital dynamics and determine whether it is captured by the Sun, i.e. if the perihelion of the hyperbolic trajectory is less than $R_{\odot}$. Particles that are determined to have a path intersecting the solar surface are then saved for the next step of the simulation. From our original analyzed sample of $2 \times 10^{10}$, we find that only approximately $3.6 \times 10^4$ particles have the initial conditions that will eventually lead to a successful capture. As expected, only a very small fraction of dark matter particles in the solar neighborhood are actually incident upon the Sun. It is important to keep in mind that this fraction does \emph{not} represent the true impact rate; calculating the true rate requires an exact measure of time duration of AQN accretion in addition to the number density rescaling. This calculation is ultimately addressed in the following sub-section (see in particular eqs. \ref{dm_scaling_factor} and \ref{impact_rate}). What \emph{is} important about these 36000 particles is that they provide us with a set of particles whose initial conditions sample exactly the true parameter space of particles captured by the Sun. Trajectory and impact properties are calculated for these particles, the distributions of which are given in figure \ref{AQN_initial_dists}.

\subsubsection{\label{2}AQN mass relations}
\begin{figure*}
\centering
\includegraphics[width=0.9\textwidth]{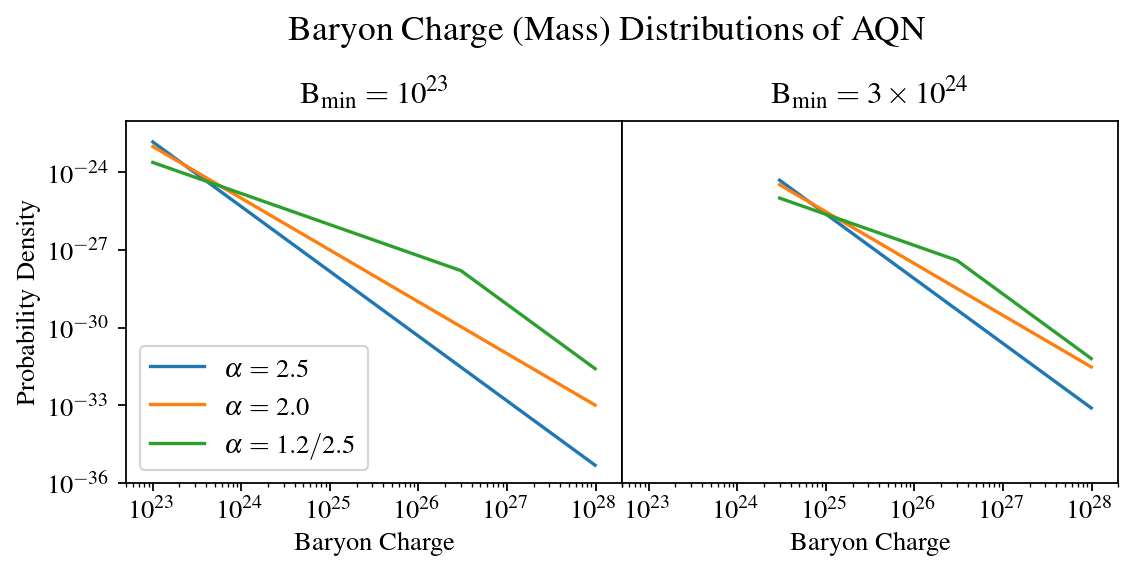}
\caption{\label{AQN_mass_dists} Probability density functions for the baryon charge B of the AQN. This directly translates to the mass distribution, as $M \simeq m_pB$.}
\end{figure*}

In order to solve the annihilation equations of the third step, and to calculate the true rate of impact events, we have  to provide the dark matter particles with a realistic mass distribution.
As discussed in section \ref{sec:QNDM}, we propose that the dark matter particles  are represented by AQNs, and the AQN annihilation events are identified with nanoflares (Sec. \ref{sec:aqn-nano}).
The direct consequence of this identification is that the nanoflare energy distribution (\ref{distribution}) coincides with the AQN mass distribution as advocated in \cite{Zhitnitsky:2017rop,Zhitnitsky:2018mav}. This identification also implies that we can adopt a variety of  models for nanoflare energy distribution which have been previously discussed in order to fit the observations. To be more specific, we use the following nanoflare models \cite{Benz-2002,Pauluhn:2006ut,Bingert:2012se} with a range of different power-law index $\alpha$ and different lower limits of extrapolation for the nanoflare energy distribution. These models have been reviewed in section \ref{sec:background}, and we plot the corresponding energy distributions in Fig. \ref{AQN_mass_dists}, where we express the energy scale in terms of the baryon charge $B$ of the AQNs according to (\ref{eq:aqn-nano}).

In our work we then explore the results of applying each of these distributions in our numerical simulations. In particular, depending on the model, the index $\alpha$ takes values of 2.5, 2.0, and a broken power-law of 1.2 below $B_{\rm threshold} \simeq 3 \times 10^{26} \ (W_{\rm threshold} \simeq 10^{24} \ {\rm erg})$, and 2.5 above. $B_{\rm min}$ is taken to be either $10^{23}$ or $3 \times 10^{24}$ ($W_{\rm min} \simeq 3 \times 10^{20} \ {\rm erg}$ or $10^{22} \ {\rm erg}$). As with the velocity, random draws of baryon charges are made from these distributions and assigned to the 36,000 impacting AQN.

As we shall see later, our main results are not very sensitive to the specific features of these different distributions. However for the purposes of our calculations, one important consequence of varying the baryon charge distribution is in determining the true number density of the dark matter particles in the solar neighborhood. This in turn determines the AQN impact rate on the solar surface and thus the proposed luminosity from these impacts. For our work, we use the current estimate for the local dark matter density of $\rho_{\rm DM} \sim 0.3~ {\rm GeV\,cm^{-3}}$. Under the AQN DM model, approximately $3/5$ of the mass density is in the form of the anti-nuggets (which are the ones that are proposed to annihilate)\cite{Lawson:2013bya}. Only some portion of this DM component ($\sim 2/3$) contributes to the annihilation processes in the solar atmosphere,  while  the remaining part ($\sim 1/3$) will be radiated as free propagating axions \cite{Ge:2017idw}. For each baryon distribution case, we can then calculate a scaling factor $f_S$ by which our results of simulating only $2 \times 10^{10}$ particles can be multiplied by to get the extrapolated true values. We have:
\beq
\label{dm_scaling_factor}
\langle B\rangle &=& \int^{10^{28}}_{B_{\rm min}} B \cdot f(B) \ dB, \qquad f(B) \propto B^{-\alpha} \qquad \qquad \nonumber \\
\bar{n}_{\rm AQ{N}} & \simeq & \left(\frac{2}{3} \cdot \frac{3}{5} \cdot 0.3~ {\rm GeV cm^{-3}}\right) \frac{1}{m_p\langle B\rangle}~~~ \qquad \qquad \nonumber \\
f_S & \equiv & \frac{\frac{4}{3} \pi (R_{\rm max}^3 - R_{\rm min}^3) \ \bar{n}_{\rm AQ{N}}}{2 \times 10^{10}},\\
\quad R_{\rm max} &=& 10 {\rm \ AU}, \quad R_{\rm min} = R_{\odot}
\eeq
The notation $\bar{n}_{\rm AQ{N}}$ is introduced to describe the true number density of the anti-matter AQNs. Put another way, the factor $f_S$ would be 1 if we populated our simulation space with the true total number of AQN in the $R_{\rm max}$ sphere instead of $2 \times 10^{10}$. The true rate of impacts of the AQN can then be approximated by considering the number of impacts $N(\Delta t_{\rm imp})$ in our sample that occur in some time window $\Delta t_{\rm imp}$, where $t_{\rm imp}$ is the time it takes for a DM particle to travel from its initial position to the solar surface. This time window cannot be chosen arbitrarily, but motivated by the finite-size effects of our simulation space. Consider the particles that initially lie on the edge of our volume, at $R_{\rm max} = 10 \ {\rm AU}$. From eq. \ref{velocity_dist}, the maximum initial velocity that these particles can have is $\sim 600 \ {\rm km\,s^{-1}}$, and if they are on a straight radial trajectory towards the Sun (the shortest path), the time it would take them to reach and impact the Sun would be $\sim 10 \ {\rm AU} / 600 \ {\rm km\,s^{-1}} \sim 1 \ {\rm month}$. Thus our time window to count the number of impacts cannot exceed $\sim 1 \ {\rm month}$. If it did, then particles that actually exist beyond 10 AU would not be correctly accounted for in our simulation space and time. This whole argument and motivation is readily apparent in the bottom-left sub-plot of figure \ref{AQN_initial_dists}, where we see as expected that the number of impacts starts to decrease beyond $t_{\rm imp} \sim 1 \ {\rm month}$, whereas before that the impact flux is constant (except for in the very beginning, where it is slightly higher due to some initial simulation effects). We thus select the precise time window which starts at $t_{\rm imp} = 0.25 \ {\rm months}$ and ends at $t_{\rm imp} = 1.25 \ {\rm months}$. Our extrapolated true impact rate calculation then follows:
\be
\label{impact_rate}
\frac{dN_{\rm imp}}{dt} \simeq \frac{N(\Delta t_{\rm imp})}{\Delta t_{\rm imp}} \cdot f_S, \quad t_{\rm imp} \in [0.25, 1.25] \ {\rm months}
\ee
For the different mass distributions that we explore, this final extrapolated impact rate varies from $\sim 10^6 ~{\rm s^{-1}}$ to $\sim 10^3 ~{\rm s^{-1}}$ for the mean baryon charges of $\langle B \rangle \sim 10^{23} - 10^{26}$ (see Fig. \ref{Tot_Extrap_Imp}).

\subsubsection{\label{3}AQN annihilation in the sun}
We now have all the dark matter parameters assigned in order to simulate the annihilation of the AQN in the solar atmosphere. Two first order differential equations have to be solved: one that describes the kinetic energy loss of the AQN due to friction as it collides with particles in the atmosphere (ram pressure), and the other that describes the mass loss of the AQN due to the annihilation of the anti-baryons of the nugget with the baryons in the atmosphere. The energy lost is assumed to radiate isotropically from the nugget surface. The equations to solve are constructed as follows:

We follow the conventional idea first formulated  by A. De Rujula and S. Glashow in a 1984 paper \cite{DeRujula:1984axn} regarding the collision of quark nuggets with the Earth.  The energy loss is:
\be
\label{energy_loss_orig}
\frac{dE}{ds} = -\sigma \rho v^2
\ee
where $s$ is the path distance, $\sigma$ is the effective cross sectional area of the nugget, $\rho$ is the density of the environment and $v$ is the nugget velocity. Re-formulating as a time derivative:
\be
\label{chain_rule}
\frac{dE}{dt} = \frac{dE}{ds} \cdot \frac{ds}{dt} =  -\sigma \rho v^3 = -\pi R_{\rm eff}^2 \rho v^3,
\ee
where we introduce the effective cross section in terms of the effective size of the nugget $R_{\rm eff}$, to be identified in what follows with $R_{\rm eff}$ from previous section.
Now, we also have:
\be
\label{kin_energy_diff}
E = \frac{1}{2} mv^2 \ \Longrightarrow \ \frac{dE}{dt} = mv \cdot \frac{dv}{dt} + \frac{1}{2} v^2 \cdot \frac{dm}{dt}
\ee
And the rate of mass loss of the AQN is given by:
\be
\label{mass_diff}
\frac{dm}{dt} = -\sigma \rho v = -\pi R_{\rm eff}^2 \rho v.
\ee
Equating eqs. (\ref{chain_rule}) (\ref{kin_energy_diff}), and substituting eq. (\ref{mass_diff}) we arrive to the following relation describing the variation of the velocity $v(t)$ of the nugget of mass $m(t)$ in the environment characterized by density $\rho$ which also varies as the nugget propagates from high latitude with low densities to lower altitude with much higher densities,
\be
\label{vel_diff}
m \cdot \frac{dv}{dt} = -\frac{\pi}{2} R_{\rm eff}^2 \rho v^2.
\ee
In vector form the complete dynamical equation of motion is then:
\be
\label{motion_vec}
m(t) \cdot \frac{d\vec{\bf v}}{dt} = -\frac{\pi}{2} R_{\rm eff}^2(t) \rho(t) v^2(t) {\bf \hat{v}} - \frac{GM_{\odot}m(t)}{r^2(t)} {\bf \hat{r}}
\ee
In order to numerically solve this equation, we must break it down into its component equations. We naturally use the circular coordinates $(r, \theta)$ (i.e. radial and tangential velocity components), and after taking into account the kinematic terms arising from our choice of coordinates, end up with the coupled differential equations:
\beq
\label{final_diff_eqns}
\frac{dv_r}{dt} &=& -\frac{v_r}{v} a - \frac{GM_{\odot}}{r^2} + \frac{v_{\theta}^2}{r}, \qquad \frac{dr}{dt} \equiv v_r \nonumber \\
\frac{dv_{\theta}}{dt} &=& -\frac{v_{\theta}}{v} a - \frac{v_rv_{\theta}}{r}, \\
\qquad \frac{dm}{dt} &=& -\frac{2ma}{v}~ \nonumber \\
{\rm with} \quad a &\equiv & \frac{\pi R_{\rm eff}^2 \rho v^2}{2m}, \quad v \equiv \sqrt[]{v_r^2 + v_{\theta}^2}~~~
\eeq
Finally, combined with the solar density and temperature profiles from Fig. \ref{observations}, and following the arguments laid out in Sec. \ref{sec:aqn-nano}, we deal with the computation of the effective radius $R_{\rm eff}$ in our numerical analysis. We treat  $m, \ v$ as the dynamical variables of the AQN, while $ \rho, \ T$ as the external parameters describing the solar atmosphere (which also depend on time $t$ as the nuggets traverse through the solar atmosphere). The resulting equations are solved for \emph{each} time step with environment dependent dimensionless parameters $\epsilon_1, \epsilon_2, M$  (as defined in Sec. \ref{sec:aqn-nano} B) calculated as follows:
\beq
\label{R_eff_calc}
M &=& \frac{v}{c} \left(\frac{3\gamma T}{m_p}\right)^{-1/2} \simeq 4.9 \left(\frac{v}{\rm km \cdot s^{-1}}\right) \left(\frac{T}{\rm K}\right)^{-1/2} \nonumber \\
\epsilon_1 &=& R \left(\frac{8m_eT}{\pi}\right)^{1/2} \ \simeq 4.7 \left(\frac{m}{\rm g}\right)^{1/3}  \left(\frac{T}{\rm K}\right)^{1/2} \nonumber \\
\epsilon_2 &=& \left(\frac{(5M^2 - 1)(M^2 + 3)}{16M^2}\right)^{3/2} \nonumber \\
R_{\rm eff} &=& \epsilon_1 \epsilon_2 R = \epsilon_1 \epsilon_2 \left(\frac{3m}{4\pi\rho_n}\right)^{1/3}, \nonumber\\
\rho_n &=& 3.5 \times 10^{17} \ {\rm kg.m}^{-3},
\eeq
where $\rho_n$ is the typical nuclear density which enters the computations  of the AQN masses. It describes the energy density per unit baryon charge.

These parameters (\ref{R_eff_calc}) play precisely the key role in our analysis as they determine the effective interaction of the AQNs with the solar material $\sigma=\pi R_{\rm eff}^2$.
This interaction obviously depends on the velocities of the AQNs because the effective coupling is proportional to the Mach number $M =v/c_s$. The effective interaction is  highly sensitive to the temperature of the environment $T$ because the ionized charge of the nugget is determined by the surrounding temperature.

We use a 4th order Runge-Kutta numerical integrator to solve the system of ODEs (\ref{final_diff_eqns}) with parameters (\ref{R_eff_calc}) determined by the environment. Looking at Fig. \ref{observations}, it is clear that the solar density is extremely low beyond a height of about 3000 km, and so there will be virtually no energy loss for the AQN before it reaches this height. Thus we start our numerical solver at a height of 3000 km for each nugget as it heads towards the solar surface. The initial radial and tangential velocities, as well as the initial masses of the nuggets are known at this height from the first two steps described in this section. The solver is allowed to run until one of two pre-defined termination events occur: i) the nugget reaches zero height i.e. hits the photosphere, or ii) the nugget loses 99.9\% of its initial mass i.e. virtually all its mass. To minimize numerical error, the maximum time step allowed is 0.01 seconds, which is on top of the in-built error tolerances of the solver, which keeps the local error estimate for $dx/dt$ below $10^{-3}x + 10^{-6}$ (for any variable $x$).

\subsection{Results}

\begin{figure*}
\centering
\includegraphics[width=\textwidth]{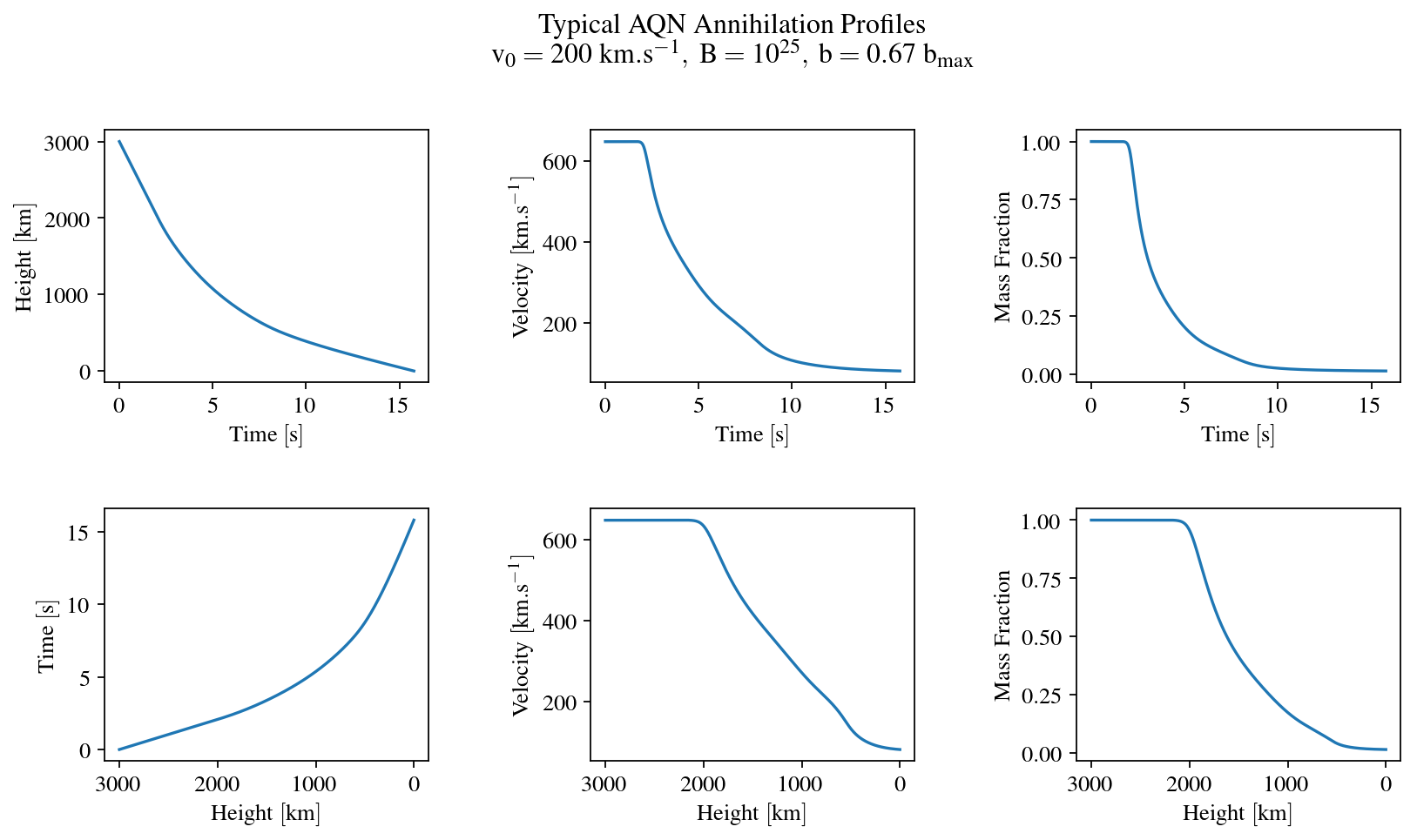}
\caption{\label{Typical_AQN_Ann_Profs} Evolution of the properties of a typical AQN as it annihilates in the solar atmosphere. In the first row the x-axis is time and in the second row it is the height. Particularly important for our proposal is the bottom-right sub-plot showing the mass lost to the environment as a function of height above the surface.}
\end{figure*}

\begin{figure}
\centering
\includegraphics[width=0.5\textwidth]{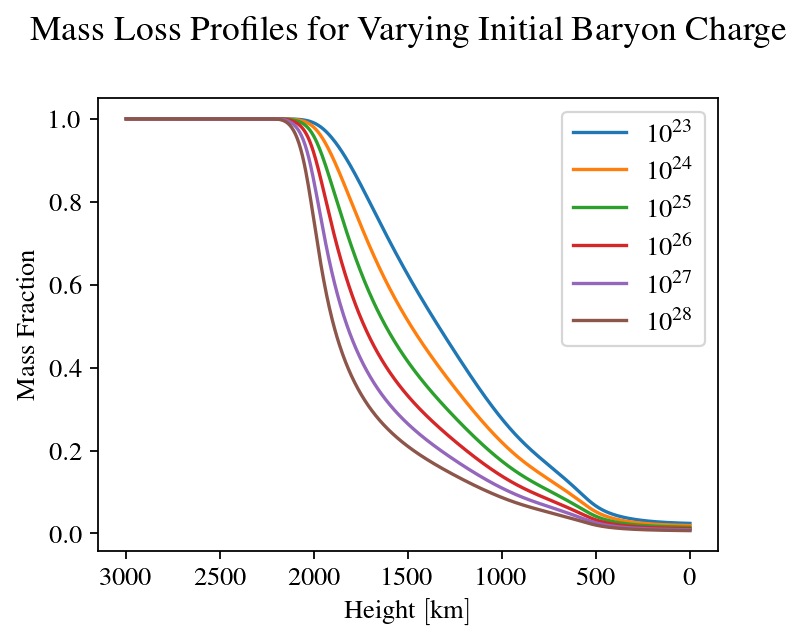}
\caption{\label{Var_Loss_Profs} The mass loss profiles for the entire range of AQN masses that we explore. It can be seen that the profile shape does not change considerably. In particular, all the AQN start to lose their mass at a height of $\sim$ 2000 km, and all have effectively lost their entire mass by the time they hit the photosphere.}
\end{figure}

$\bullet$ The time (and height) dependent solution for a typical AQN trajectory as it annihilates in the solar atmosphere is shown in figure \ref{Typical_AQN_Ann_Profs}.  The same parameters as a function of the height above the photosphere are shown at the bottom row in figure \ref{Typical_AQN_Ann_Profs}.  There are two key observations here.
The first is  that the nugget loses virtually all its mass before reaching the photosphere as shown on the   bottom right panel figure \ref{Typical_AQN_Ann_Profs}, thus confirming the original assumption \cite{Zhitnitsky:2017rop} and fully consistent with our  present proposal. Furthermore, we find an even more profound  feature: the AQN starts to lose energy to the environment at a height of about 2000 km, which is where the solar Transition Region is. What is most remarkable about this feature is that it is a very robust property of the system, and not very sensitive to the specific details of the model. Indeed, if  we vary  the  masses of the AQNs, we still get the same starting height
around 2000 km, and similar profiles overall, as seen in figure \ref{Var_Loss_Profs}.

\begin{figure*}
\centering
\includegraphics[width=\textwidth]{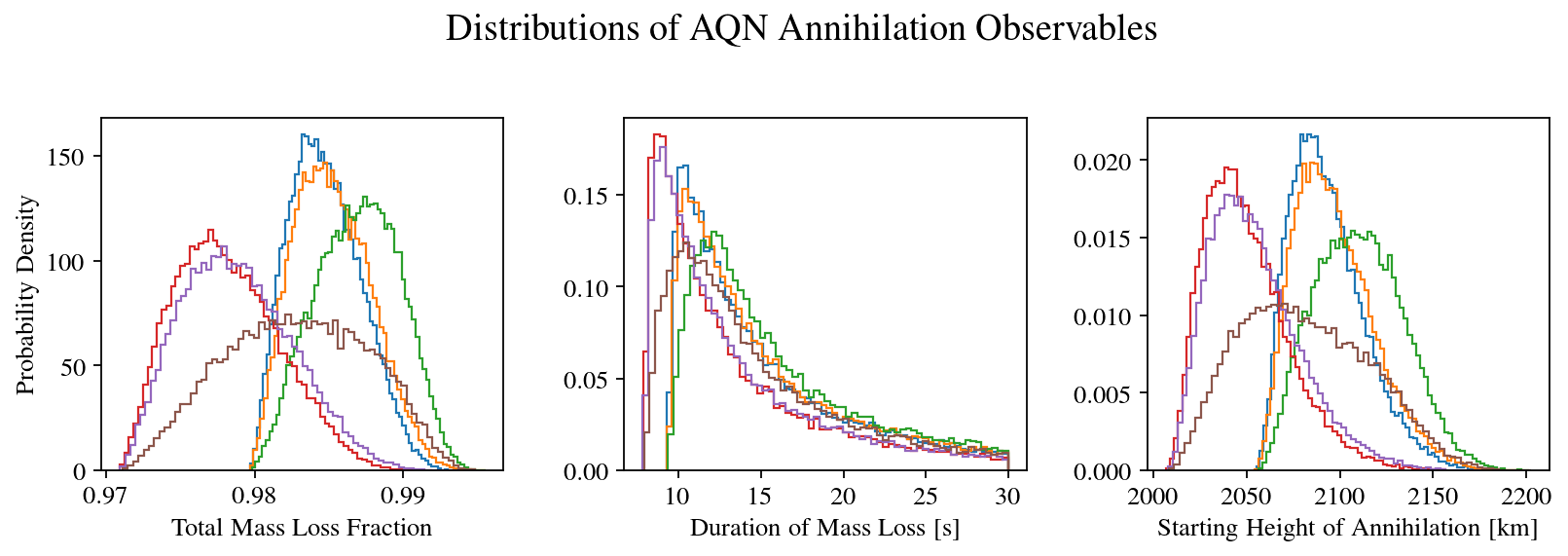}
\caption{\label{Observ_Dists} The distributions of some annihilation observables that we get from simulating the evolution of all AQN as they travel through the solar atmosphere. The different colors correspond to the different mass distributions we explore as given in figure \ref{AQN_mass_dists}. What is important is not the slight differences between distributions, but the narrow range of values that the results cover over the entire parameter space.}
\end{figure*}

\begin{figure}
\centering
\includegraphics[width=0.5\textwidth]{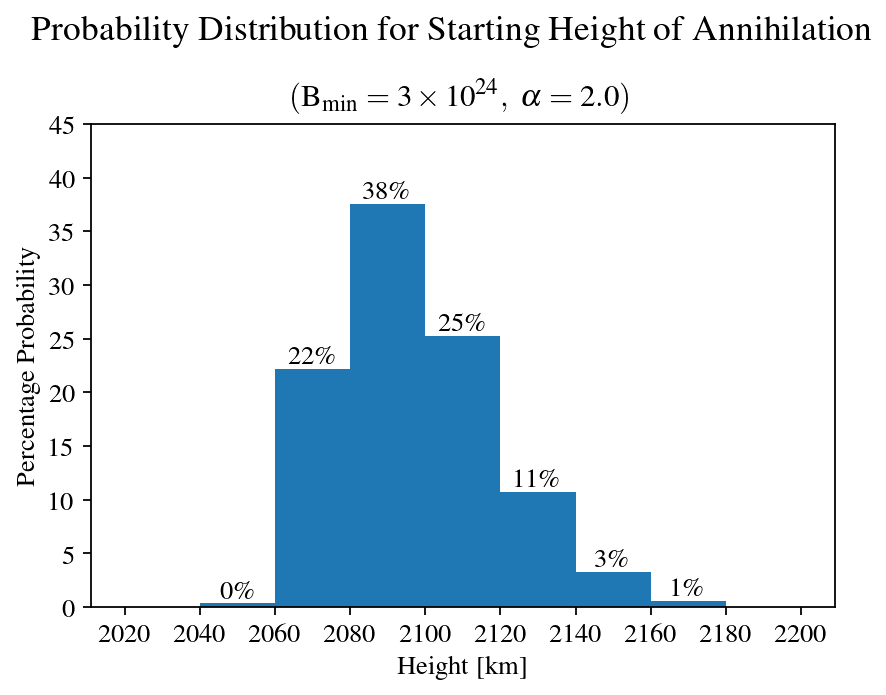}
\caption{\label{Height_Percentage} An easy to read plot for the distribution of the annihilation starting height, which is defined to be the altitude at which the AQN has lost 0.5\% of its initial mass.}
\end{figure}

\bigskip
\noindent $\bullet$ We then check whether these features are indeed very robust consequences of the entire system, not being too sensitive to the details of the nanoflare energy distributions listed in figure \ref{AQN_mass_dists}, nor to the range of initial conditions for the impacting AQN.
With this goal in mind, we solve for the evolution of the AQN in the solar atmosphere for all the $\sim$ 36000 DM particles, and repeat this exercise for the 6 different mass distribution models. The results of this analysis is shown in figure \ref{Observ_Dists}. Three important features stand out:

a) All nuggets, regardless of initial mass, velocity or impact parameter, annihilate and lose more than 97\% of their total mass to the environment before hitting the surface;

b) The timescale for this loss is on the order of $10$ seconds, which is completely consistent with the time-scales expected for nanoflare events (of order $10^1 - 10^2$ s). For our analysis we have defined the annihilation starting time to be when the AQN has lost 0.5\% of its initial mass, and the ending time to be when it hits the surface (or has only 0.5\% of its initial mass left);

c) All nuggets start annihilating between 2000-2200 km, almost exactly overlapping with the Transition Region.
The same feature can be represented in a different way by plotting the probability distribution in percentage as a function of the height, shown in figure \ref{Height_Percentage}. For illustrative purposes we only show a particular mass distribution, but this generic feature holds for other distributions as well.

\begin{figure}
\centering
\includegraphics[width=0.5\textwidth]{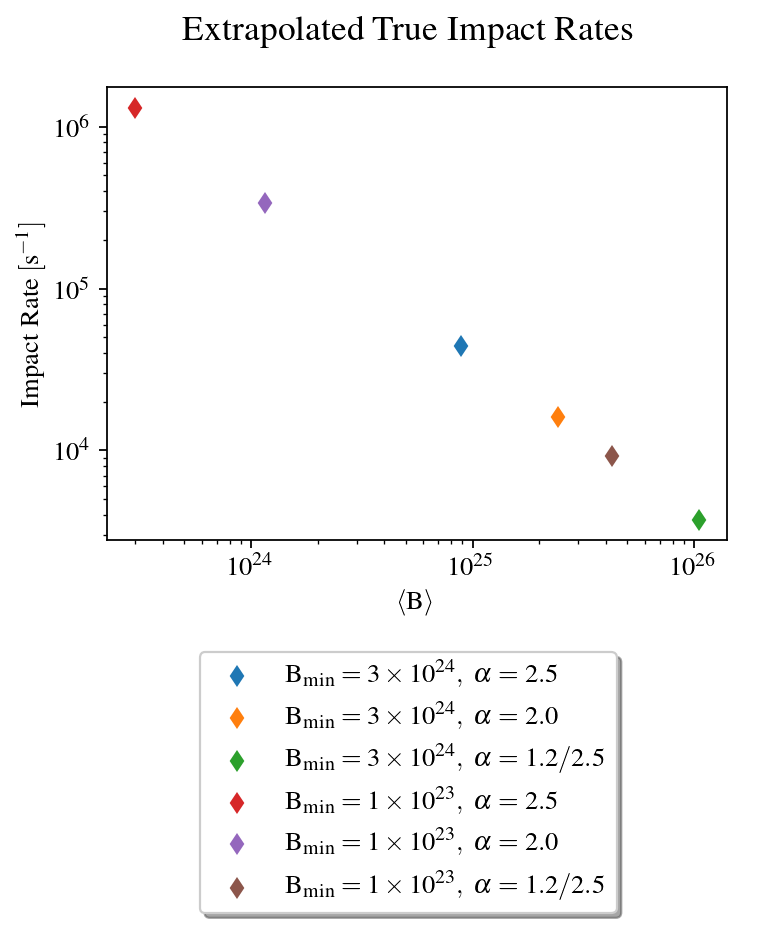}
\caption{\label{Tot_Extrap_Imp} The extrapolated true impact rate for the different mass distributions presented in figure \ref{AQN_mass_dists}, and calculated according to eq. (\ref{impact_rate}). In the framework of our proposal, this is the same as the nanoflare event frequency.}
\end{figure}

\bigskip
\noindent $\bullet$ The next important result is that of the AQN impact rate, which is by definition the number of annihilation events per second that happen over the entire solar surface as a result of AQNs impacting the Sun (see eq. \ref{impact_rate}). The corresponding plot is presented in figure \ref{Tot_Extrap_Imp}. The rate depends on the AQN mass distribution model, which are shown in figure \ref{AQN_mass_dists}. As shown in figure \ref{Tot_Extrap_Imp}, the less massive the nugget, the higher the impact rate.

\begin{figure}
\centering
\includegraphics[width=0.5\textwidth]{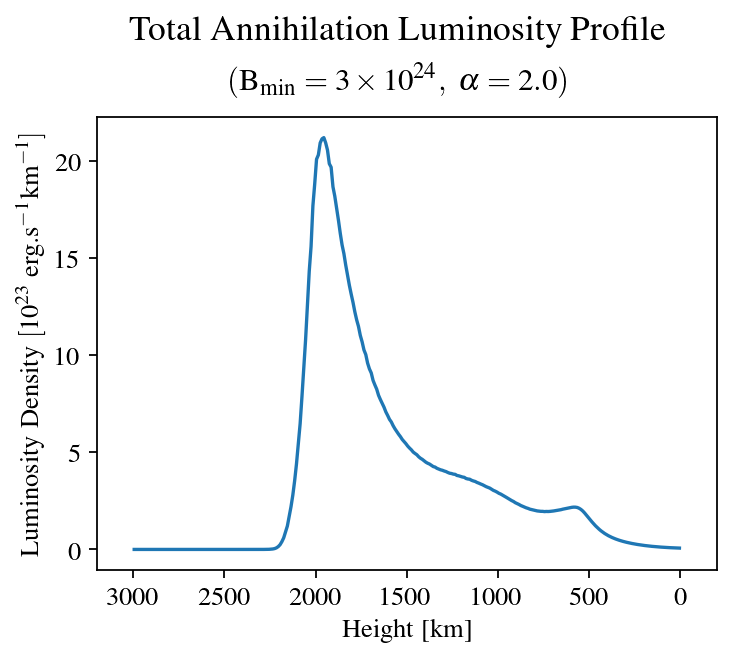}
\caption{\label{Tot_Lum_Density} The total deposited  energy profile for a particular mass distribution. Here the \emph{total} energy injection is calculated by multiplying the mean annihilation energy profile for the AQNs by the extrapolated total impact rate. The luminosity peak seen at $\sim$2000 km serves to suggest a natural explanation within our model for the temperature rise in the Transition Region.}
\end{figure}

\bigskip
\noindent $\bullet$ Following the previous result, we want to compute the total injected energy per unit time per unit length for a given altitude over the entire solar surface. We want to quantify the AQN energy deposition as a function of height (i.e. the annihilation luminosity density). The result is presented in figure \ref{Tot_Lum_Density} for a typical nanoflare distribution. The corresponding behavior is striking: it is strongly peaked at a height around 2000 km, in close vicinity of the Transition Region. This profile shape is robust and holds for all nanoflare distributions listed in figure \ref{AQN_mass_dists}.

The technical reason for this behaviour to emerge is related to the drastic changes that occur in the interaction rate of the AQN with the solar material. The corresponding effective interaction cross section depends on the temperature, density and the Mach number, and all these parameters rise (or fall) sharply in the Transition Region. We speculate that this local very fast and efficient deposition of energy is a key element in solving the Transition Region puzzle with its dramatic variation of all thermodynamical  parameters on a small scale (measured in $10^2$~km rather than in $10^3$~km), as shown in figure \ref{observations}.

\begin{figure}
\centering
\includegraphics[width=0.5\textwidth]{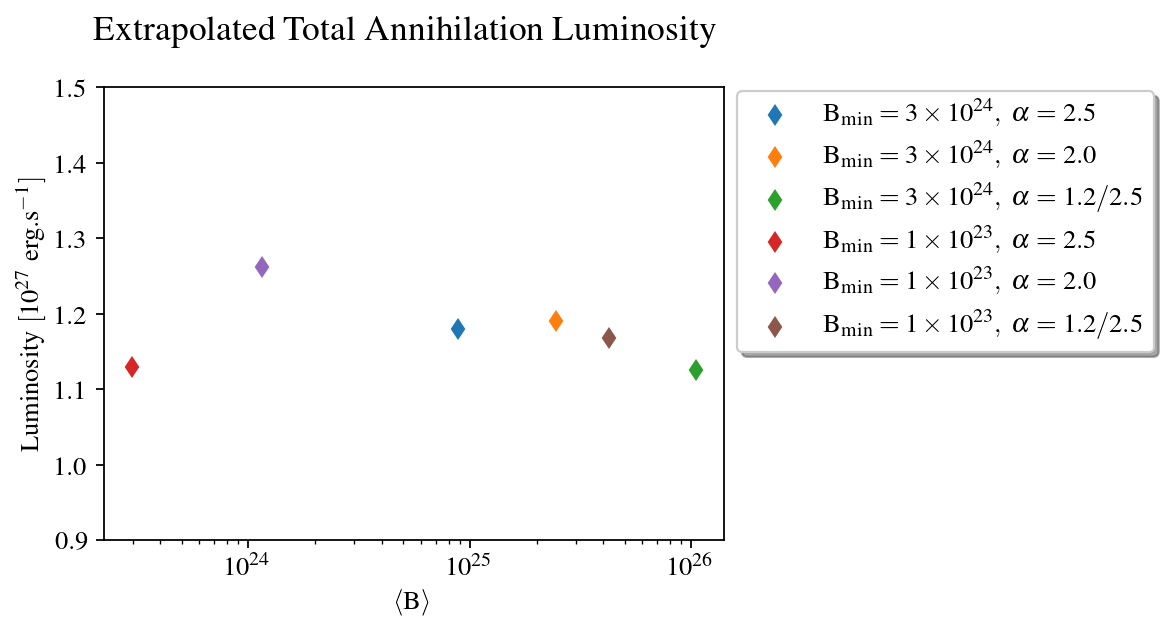}
\caption{\label{Tot_Extrap_Lum} The extrapolated total luminosity for the different mass distributions, that are a result of our simulations (following color scheme of Fig. \ref{Tot_Extrap_Imp}). Remarkably, without any fine-tuning of parameters, the total luminosity is about $1 \times 10^{27} \ {\rm erg~s^{-1}}$ across distributions, in agreement with the observed quiet Sun EUV and soft X-ray flux.}
\end{figure}

\bigskip
\noindent $\bullet$ Our final comment relates to the computational result for the total annihilation energy injected in the solar atmosphere per unit time (the annihilation luminosity). It is calculated as $L_{\rm tot} = \langle \Delta m_{\rm AQN} \rangle \cdot dN_{\rm imp}/dt$, where $\Delta m_{\rm AQN}$ is the total mass lost by an AQN in its trajectory through the solar atmosphere. Essentially it represents the integral over the energy distribution as a function of height shown in figure \ref{Tot_Lum_Density} (indeed both methods are self-consistent). The result of this calculation for 6 different nanoflare distributions is shown in figure \ref{Tot_Extrap_Lum}. The most profound feature of this plot is that the total luminosity (energy injection) is almost constant, and is not sensitive to the nanoflare models. Furthermore, it is amazingly close to the observed luminosity $\sim 10^{27} \ {\rm erg~s}^{-1}$ in EUV and soft X-rays radiation. This ``numerical coincidence'' was, in fact, the main motivation in \cite{Zhitnitsky:2017rop} to advocate this proposal.

 The intuitive explanation that the total luminosity is not sensitive to the AQN mass (nanoflare) distribution can be understood from the fact that our basic normalization is determined by the dark matter density $0.3 \ {\rm GeV\,cm^{-3}}$. Different distributions would generate different \emph{number} densities (and impact rates) of the nuggets as shown in Fig. \ref{Tot_Extrap_Imp}. However, the total \emph{mass} available in the solar neighborhood to annihilate is fixed, and since we have already shown that the individual mass loss fraction for nuggets is not particularly sensitive to the initial mass distribution, thus the total injected annihilation energy remains (almost) the same as well.

 Figure \ref{Tot_Extrap_Lum} also demonstrates the self-consistency of our numerical computational scheme. Indeed, we started with a very large number of particles distributed over a $10 \ {\rm AU}$ radius sphere. Nevertheless, we ended up (after a large number of pure computational steps, not related with the underlying  physics of the AQN dark matter proposal) with proper number of AQNs entering the solar atmosphere and generating the luminosity  of order $\sim 10^{27} \ {\rm erg ~s}^{-1}$.

 This is a remarkable result because this energy is mostly emitted from the region around 2000 km (as shown in figure
\ref{Tot_Lum_Density}) which is characterized by a high temperature $T\sim 10^6~$K. Therefore, it is quite  natural to expect that most of the emission will be in the form of EUV and soft X-rays, in full agreement with observations.   These results provide strong numerical support for the assumption made in \cite{Zhitnitsky:2017rop} that the  luminosity generated by the AQN annihilation events will be mostly radiated in the EUV and soft X-ray bands of the spectrum.

\section{Conclusion}\label{conclusion}


We have shown that the AQN dark matter model could account for a significant fraction, if not all, of the EUV radiation excess of the solar corona. One should emphasize that we are dealing with the quiet Sun only. In order to work in the quiet Sun regions, conventional heating models require the existence of unobserved (i.e. unresolved with current instrumentation) ``nanoflares'' \cite{De-Moortel-2015}, described as a generic small scale source of energy, the physical nature of which is unspecified \cite{Klimchuk:2005nx,Klimchuk:2017}.
Our proposal identifies the nanoflares with the AQN annihilation events \cite{Zhitnitsky:2017rop,Zhitnitsky:2018mav}.
The AQN model was initially designed to address cosmological issues expressed by Eq. (\ref{Omega}) and has no tuning parameters associated with the physics of the Sun \footnote{In particular, the model has only one tunable parameter, the axion mass scale $m_a$, since the baryon charge $B$ of the nuggets is determined by the axion mass (as reviewed in Sec. \ref{sec:QNDM})}; its consequence for the corona heating cannot be adjusted.
The main results of our numerical simulations are  expressed by two plots. First, Fig. \ref{Tot_Lum_Density} shows that the dominant portion of the energy is injected in the region close to 2000 km where $T\simeq 10^6~K$, and therefore the   radiation is expected to be in form of the EUV and soft X-rays.  Secondly, Fig. \ref{Tot_Extrap_Lum} shows that this EUV radiation is very close to the observed value $\sim  10^{27} \ {\rm erg~s^{-1}}$, which is a very direct consequence of the model mostly determined by the dark matter density $0.3 \ {\rm GeV\,cm^{-3}}$ in the Solar System.

Our study provides an energy injection scheme for the solar corona. However, the corona is optically thin, and therefore all photons created by the annihilation of AQNs should in principle escape the Sun. The process by which this energy can be re-injected in the plasma requires magnetohydrodynamics (MHD) simulations because it involves the computation of the electromagnetic interaction of the plasma with the moving, highly ionized, AQN. This is a complicated process which is left for future work. Nevertheless, our current results serve to illustrate and provide important clues that the energy injection will happen at the correct altitude, within the expected transition region and with the correct energy. 


In view of the developed framework, we also offer the following items as possible tests of our proposal:

1- The `nanoflare' sites will be observed as bursts of energy, but in the AQN model they should not be associated with any local magnetic activity (like flares do), including in quiet regions during the solar minima. Since the source of energy injection comes from a random direction from space, it can happen anywhere and not specifically in active regions where the magnetic field  is  strong.

2- The energy injection should be confined to the top of the chromosphere, in the Transition Region. It is sometimes advocated that the top chromosphere high temperature is a problem even more serious than the ``hot" corona because the chromosphere is much denser and therefore harder to heat. In our model the heating of the chromosphere should arise naturally from energy injection at this height, as one can see from Fig. \ref{Tot_Lum_Density} describing the altitude distribution of the energy deposition.

3- The altitude of energy injection should be the same everywhere, whether or not we are looking above a quiet or active region of the Sun. It is still not understood how prominences can form in the much hotter and rarefied coronal regions. In our model, prominences and coronal heating are completely different phenomena.

4- The energy injection in our model can be thought as a local event which lasts about 10 seconds with typical linear spatial extensions of order 1000 km. Conventional MHD should be used to describe consequent  temporal and spatial evolution of these energetic disturbances which should be treated as the  initial configurations of the system.

5- It is possible that high resolution imaging could reveal shock wave fronts caused by the AQNs moving at velocities much larger than the speed of sound. The observation of  these small jet-like events with typical nanoflare energies, lasting for about 10 seconds outside the active regions will be strong evidence supporting our proposal.

6- As we mentioned in section \ref{sec:QNDM} a finite portion (about 1/3 of its mass) of the AQNs will be disintegrated in the form of the propagating axions. Therefore, the total intensity of the emitted axions can be estimated as   1/2 of the EUV emission computed in this work (and plotted on Fig. \ref{Tot_Extrap_Lum}), i.e. $ 0.5 \times 10^{27} \ {\rm erg~s^{-1}}$.
These   axions will be mostly emitted with relativistic velocities $v\sim 0.5 ~c$. Therefore, they will  have  very distinct spectral properties in comparison to  galactic axions (characterized by $v\sim 10^{-3}~c$) and conventional solar axions which are produced through the Primakoff effect in the central regions of the Sun. 
This new type of the solar axions can, in principle, be discovered with upgraded CAST (CERN Axion Search Telescope) type instruments, as argued in \cite{axion}.
This new solar axion production mechanism can be tested  with upgraded CAST (CERN Axion Search Telescope) type instruments \cite{axion}. Furthermore, a similar production mechanism also generates
the axions on Earth, a fraction of which have small velocities below the escape velocity. These axions will be accumulated by the Earth (and similarly the Sun)  during its Gyr life-times, which  greatly enhances the discovery potential \cite{Liang:2018ecs}.

As our final remark, we note that NASA has recently launched a mission in August 2018, the Parker Solar Probe (PSP) \footnote{https://www.nasa.gov/content/goddard/parker-solar-probe}, designed to explore the solar corona and its heating mechanisms. PSP will be capable of performing high resolution imaging of the lower solar atmosphere and detailed studies of its magnetic environment. Thus, it should be able to address some of the tests above, with first results expected in early 2019.

\section*{Acknowledgements}

This research was supported in part by the Natural Sciences and Engineering
Research Council of Canada. We would like to thank Alex Mead, Gary Mamon and Benoite Pfeiffer for useful conversations on some aspects of this work.

       \bibliography{solar_corona}

\
    \end{document}